\documentclass[english]{scrartcl}
\pdfoutput=1

\usepackage[utf8]{inputenc}

\usepackage{graphicx}
\usepackage{%
    amsmath,
    amssymb,
    babel,
    booktabs, 
    csquotes,
    color, 
    enumitem, 
    geometry, 
    graphicx, 
    lineno,
    multirow,
    natbib,
    siunitx,
    tabularx,
    soul
       }

\usepackage[dvipsnames]{xcolor}

\usepackage{authblk}

\usepackage{amssymb}
\usepackage{amsmath}
\usepackage{bm}
\usepackage{blindtext}
\usepackage{booktabs}
\usepackage{graphicx}
\usepackage{mathtools}
\usepackage{xspace}
\usepackage[dvipsnames]{xcolor}
\usepackage{url}
\graphicspath{{fig/}}

\newcommand{\setobs}[1]{\ensuremath{\left\{#1_i\right\}}}
\newcommand{\settime}[1]{\ensuremath{\left\{t_{\uppercase{#1},i}\right\}}}
\newcommand{\obs}[1]{\ensuremath{#1_i}}
\newcommand{\obst}[1]{\ensuremath{t_{\uppercase{#1},i}}}
\newcommand{\nrho}{\ensuremath{\rho_{MCMC}\left(\mathbf{X},\mathbf{Y}\right)}}

\newcommand{\X}{\ensuremath{\mathbf{X}}\xspace}
\newcommand{\Y}{\ensuremath{\mathbf{Y}}\xspace}
\newcommand{\dO}{\ensuremath{\delta^{18}}O\xspace}

\title{Correlating Paleoclimate Time Series: Sources of Uncertainty and Potential Pitfalls}

\begin{document}
%\linenumbers

\author[1, 2]{Jasper G. Franke}
\author[1,3]{Reik V. Donner}
\date{}

\affil[1]{Potsdam Institute for Climate Impact Research, Potsdam, Germany}
\affil[2]{Humboldt University of Berlin, Berlin, Germany}
\affil[3]{Magdeburg-Stendal University of Applied Sciences, Magdeburg, Germany}

\maketitle
%\ead[em1]{jasper.franke@pik-potsdam.de}

%%% Title, authors and addresses

%%% use the tnoteref command within \title for footnotes;
%%% use the tnotetext command for theassociated footnote;
%%% use the fnref command within \author or \address for footnotes;
%%% use the fntext command for theassociated footnote;
%%% use the corref command within \author for corresponding author footnotes;
%%% use the cortext command for theassociated footnote;
%%% use the ead command for the email address,
%%% and the form \ead[url] for the home page:
%%% \title{Title\tnoteref{label1}}
%%% \tnotetext[label1]{}
%%% \author{Name\corref{cor1}\fnref{label2}}
%%% \ead{email address}
%%% \ead[url]{home page}
%%% \fntext[label2]{}
%%% \cortext[cor1]{}
%%% \address{Address\fnref{label3}}
%%% \fntext[label3]{}

%\address{}

\begin{abstract}
Comparing paleoclimate time series is complicated by a variety of typical features, including irregular sampling, age model uncertainty (e.g.,\ errors due to interpolation between radiocarbon sampling points) and time uncertainty (uncertainty in calibration), which---taken together---result in unequal and uncertain observation times of the individual time series to be correlated. Several methods have been proposed to approximate the joint probability distribution needed to estimate correlations, most of which rely either on interpolation or temporal downsampling.

Here, we compare the performance of some popular approximation methods using synthetic data resembling common properties of real world marine sediment records. Correlations are determined by estimating the parameters of a bivariate Gaussian model from the data using Markov Chain Monte Carlo sampling. We complement our pseudoproxy experiments by applying the same methodology to a pair of marine benthic \dO records from the Atlantic Ocean.

We find that methods based upon interpolation yield better results in terms of precision and accuracy than those which reduce the number of observations. In all cases, the specific characteristics of the studied time series are, however, more important than the choice of a particular interpolation method. Relevant features include the number of observations, the persistence of each record, and the imposed coupling strength between the paired series. In most of our pseudoproxy experiments, uncertainty in observation times introduces less additional uncertainty than unequal sampling and errors in observation times do. Thus, it can be reasonable to rely on published time scales as long as calibration uncertainties are not known.
\end{abstract}

%\begin{keyword}
%%% keywords here, in the form: keyword \sep keyword

%%% PACS codes here, in the form: \PACS code \sep code

%%% MSC codes here, in the form: \MSC code \sep code
%%% or \MSC[2008] code \sep code (2000 is the default)

%\end{keyword}

\section{Introduction}
While paleoclimate proxies reflect the environmental conditions at their specific location, similarities and differences among them can inform about past large scale dynamics of the climate system. Often, such a comparison is done by eye alone \citep[e.g., in ][]{cheng_climatic_2012, zhang_test_2008, waelbroeck_timing_2011}. This is, however, a highly subjective approach, as one often tends to focus on similarities at long time scales, while ignoring short time variability. As the number of high-resolution, spatially disperse paleoclimate data steadily increases, a more quantitative analysis with less ambiguity becomes both possible and necessary. 

One way to compare time series is to estimate their mutual correlation as a measure of similarity or association (we will use these three terms synonymously in the remainder of this paper). Correlations are not only interesting in themselves, they also form the basis of many analysis methods like empirical orthogonal functions \citep[EOFs,][]{mann_global-scale_1998}, complex network approaches \citep{rehfeld_late_2013, nao_paper} and climate field reconstruction methods \citep{werner_spatio-temporal_2018}. All these methods can serve as the basis of multi-proxy based index and field reconstructions of large-scale climate and/or environmental conditions in the past and have thus been central to our understanding of past climate variability. This is particularly the case for regional, hemispheric and global reconstructions of temperatures, as have been provided by, among others, \citet{mann_global-scale_1998, pages_2k_consortium_continental-scale_2013, marcott_reconstruction_2013} (to mention just a few key studies), all of which have relied on some estimate of the covariance matrix among a set of records. It is for this reason, that the problem of estimating and/or reconstructing the covariance between paleoclimate records has often been discussed in the past, mostly in the context of reconstructions of past climate indices or fields, e.g., by \citet{christiansen_challenges_2017, tingley_piecing_2012, smerdon_reconstructing_2016} and references therein. These discussions have rarely systematically considered the effect of time uncertainties, but rather focused on issues like variable record numbers and signal-to-noise ratios. They have also largely restricted themselves to well dated archives from the last millennia, in particular tree rings, varved lake sediments and ice cores, but rarely addressed records that go beyond the time scale up to which reliable dating is possible, i.e., beyond which proxy records often face more severe problems regarding their time uncertainties. 

While most similarity measures rely on concurrent observations, two common features of paleoclimate records covering earlier times, irregular sampling and time uncertainty, make the estimation difficult, as both lead to unequal observation times among time series. An additional challenge is the generally low temporal resolution of the records. Several methods have been proposed to approximate the joint distribution from unequally sampled data, often by some kind of interpolation or smoothing \citep{mudelsee_climate_2010}. Some of these methods have been compared in previous studies \citep{rehfeld_comparison_2011}, also taking time uncertainties into account \citep{rehfeld_similarity_2014}.

Here, we use a Bayesian approach to estimate the association between the time series instead of applying classical point estimation. This procedure is advantageous to classical approaches in many ways. First of all, it makes the assumptions of the underlying statistical model explicit and might thus prevent application in inappropriate situations, for example, for data that deviate strongly from a Gaussian distribution. With the use of appropriate prior distributions \citep{lewandowski_generating_2009}, it is possible to ensure that the estimated correlation matrix is positive semidefinite, a feature that is not guaranteed by classical estimators when interpolation is applied \citep{rehfeld_comparison_2011, babu_spectral_2010}. This characteristic is central for many statistical methods that built upon the covariance matrix. For example, a non positive-semidefinite covariance matrix has negative eigenvalues and therefore, the interpretation of explained variance by the corresponding EOF analysis is not feasable anymore.

In addition, the posterior distributions of parameters give us information about the estimation uncertainty, e.g., due to a low number of observations. Similar information has otherwise to be obtained using the Fisher z-transform as an approximation of the sample distribution \citep{fisher_frequency_1915}. In the presence of measurement uncertainty the Bayesian approach has been reported \citep{behseta_bayesian_2009, matzke_bayesian_2017} to be superior to classical corrections based on the work of \citet{spearman_proof_1904}. As the issue of observation uncertainty has been discussed in more detail in the aforementioned publications, we focus here on the effects that irregular sampling and age models have on the estimation of correlation. 

Hence, in this study we aim to understand how a probabilistic framework can lead to more meaningful estimates of correlation by incorporating intrinsic uncertainties typical for paleoclimate records. The contributions of different levels of time uncertainties to the overall estimation error and uncertainty are compared. With this we attempt to provide recommendations for future work that relies on Bayesian correlation estimation of paleoclimate data. 

The remainder of this paper is organized as follows: In Sec.~\ref{sec:methods} we discuss the Bayesian framework for estimating correlations among time series and different approximation techniques. We furthermore introduce the synthetic time series used in this study, resembling marine sediment records with their associated uncertainties from varying sedimentation rates and radiocarbon dating. In Sec.~\ref{sec:results} we study, how different parameter values of the pseudoproxies influence the estimation of correlation and lags. These techniques are then applied to a pair of marine records in Sec.~\ref{sec:real-world}. We end with some concluding remarks about what these results imply for comparing any pair of paleoclimate time series.

\section{Methods}
\label{sec:methods}

\subsection{Similarity measures}
We are interested in comparing two time series $\mathbf{X}$ and $\mathbf{Y}$. A time series $\mathbf{X}$ is a set $\left\{\obs{x},\obst{x}\right\}$, with observations $\vec{x}=\setobs{x}$ and observation times $\vec{t}_X = \settime{x}$, $i=1,2,\dots,N_x$, with $N_x$ being the number of observations. The observations of a time series are drawn from a distribution $P(x)$. 
If $\settime{x} = \settime{y}$ holds for the two time series, then a joint probability distribution $P(x,y)$ exists. The two sets of observations are statistically independent if and only if $P(x,y) = P(x)P(y)$,  otherwise they are said to be dependent. A common way to describe the structure and strength of such a dependence is to employ certain measures quantifying the statistical association between two sets of observations. In general, such a measure is expected to be zero if $\mathbf{X}$ and $\mathbf{Y}$ are independent and non-zero otherwise, reaching its maximum value if both are identical.

Some methods, like mutual information \citep{paninski_estimation_2003}, operate on the joint and marginal probability distributions directly. They face the problem that the estimation from limited, uncertain data is non-trivial and can introduce biases \citep{paninski_estimation_2003, papana_evaluation_2009}. In these situations it can thus be a better idea to use simple statistical models as approximations, whose parameters can be estimated more precisely.

A popular model used to describe $P(x,y)$ is the bivariate normal model (BNM)
        \begin{equation}
         P(x,y) \sim \mathcal{N}\left(\vec{\mu},\underline{\Sigma}\right)\label{eq:normal1} 
     \end{equation}
with 
     \begin{equation}
\vec{\mu} = \left\{\langle x\rangle,\langle y\rangle \right\} \quad\text{and}\quad \underline{\Sigma} = \begin{pmatrix}            \sigma_X^2 & \rho\sigma_X\sigma_Y \\ \rho\sigma_X\sigma_Y & \sigma_Y^2\end{pmatrix}.
       \label{eq:normal} 
     \end{equation}
Here, $\sigma_X^2$ and $\sigma_Y^2$ are the respective variances of the corresponding time series \X and \Y, while $\rho$ describes the shared variability (i.e., their linear correlation or  coupling strength). This poses a linear model of statistical association, as the full interdependence is described by a linear term in the covariance matrix. A common estimator of the correlation coefficient $\rho$ is the Pearson correlation, which, for sufficiently long time series, is asymptotically unbiased and efficient \citep{lehmann_theory_1998}.

In this work, we will follow the above idea and restrict our analysis to the problem of estimating the parameters of this BNM. We consider the parameter $\rho$ as the measure of similarity, also referred to as correlation. 

To estimate the model parameters, we use a Markov Chain Monte Carlo (MCMC) approach with Metropolis sampling \citep[for an overview of Bayesian methods and MCMC, see, e.g.,\  ][]{von_toussaint_bayesian_2011, gelman_bayesian_2014}. As with all Bayesian approaches, we need to define prior probability distributions that represent our knowledge, before estimation. These will be denoted as $P_{pr}(X)$ for any variable $X$. As we like to make our analysis as general as possible, we use very weakly informed priors. This means, that we impose distributions, but with very broad variances. All the priors assume that each time series has been normalized to zero mean and unit variance before estimation. Then, the priors for the mean $\mu$ and variances $\sigma$ are chosen as  $P_{pr}(\mu) \sim \mathcal{N}(0, 10)$ and $P_{pr}(\sigma) = \text{HalfCauchy}(2.5)$ \citep[for a discussion of this distribution see, ][]{polson_half-cauchy_2012}.
In most situations, the positive semidefinitive nature of the covariance matrix is the only prior information we have. \citet{lewandowski_generating_2009} describe a prior to efficiently sample from the set of positive semidefinite matrices, the so called LKJ prior. It introduces an additional parameter $\eta$, for which we set a uniform prior of $P_{pr}(\eta) = \text{Uniform}(0, 5)$.
We draw 30,000 samples and discard the first third of them, as the sampler has not converged for these samples. For these settings, we did not detect any non-convergence for time series as discussed in this study. To save memory and storage, we keep only 1,000 values for each pair of time series.

The result of the MCMC estimation are posterior probability distributions for the model parameters. The only one we are interested in here is the posterior distribution of the coupling parameter $\nrho\sim p\left(\left\{\rho , i=1,\dots, N_{\text{steps}}\right\}\right)$, with $N_{\text{steps}}$ denoting the number of steps of the MCMC sampler.  The corresponding point estimate $\hat{\rho}\left(\mathbf{X},\mathbf{Y}\right)$ is given by the value of highest probability in $\nrho$.

\subsection{Irregular sampling and age model uncertainties}
\label{sec:irregular}
Apart from measurement uncertainty, two problems make the estimation of correlations difficult in the case of paleoclimate records: irregular sampling and age model uncertainty.

A time series is \emph{irregularly sampled} if $\Delta t_{X,i+1|i} \coloneqq t_{X,i+1}-t_{X,i}$ is different for different $i$. Many paleoclimate archives have varying recording conditions (e.g.,\ sedimentation rate) and, thus, even regular physical sampling intervals (e.g.,\ equally spaced depths along a sediment core) will usually lead to irregular sampling in time. 

When comparing two irregularly sampled time series $\mathbf{X}$ and $\mathbf{Y}$, they are commonly \textit{unequally sampled}, meaning that the observation times are unequal, $\settime{x} \neq \settime{y}$. In this case, $P(x,y)$ cannot be estimated directly and has to be approximated.

Age models add another level of uncertainty to the estimation. Only a few paleoclimate archives, like tree ring, show a reliable relative or absolute time scale. In most cases, the attribution of times to observations is done in indirect ways, e.g.,\ by the radioactive decay of isotopes like $^{14}$C \citep[see, e.g.,\ ][]{bradley_paleoclimatology_2015}. In this study we distinguish between two sources of error arising from age models. The first is related to the fact that all age models make assumptions about the archive evolution (like constant sedimentation rate in the most simplest models that use linear interpolation between age control points (e.g., the program clam by \citet{blaauw_methods_2010}) between points of age measurements. Thus, most points are subject to error which is hard to quantify. Second, each observation is not attributed to one point in time, but rather to a set of possible times. This uncertainty we will address mainly by using different realizations of the age model for our estimation.

Since concurrent observations are an exception for pairs of paleoclimate records, it is necessary to approximate the joint distributions from the marginal distributions. In this study, we focus on three possible approaches to tackle this problem: (i) interpolation of both variables to a shared time axis, (ii) comparing averages over time intervals, and (iii) combining observations which are presumed to be close to each other. Specifically, the following four methods will be used in this study:
\begin{description}
    \item[Linear interpolation (LI).] Here, missing observations are assumed to lie on a straight line between the two neighbouring observations and, hence, the point on this imaginary line corresponding to the specific time is set as a value.
    \item[Gaussian kernel interpolation (G).] Here, the unobserved value is set to a weighted mean of nearby observations in time, with close points having larger weights than those that are farther away in time. How many observations are taken into account depends on the bandwidth $(h)$ of the weight function and has to be specified in advance. As we use different bandwidths, we denote the corresponding approximation as G(bandwidth), with the bandwidth values being scaled by the mean sampling time of a given pair of time series.
    \item[Nearest value (NV).] In this approach, values that are close to each other in time are considered to be concurrent, whereas all others are discarded, hence lowering the effective number of observations, but also not introducing additional data points.
    \item[Slotting (S).] Values are averaged over time slots for both time series and these mean values are then compared. This is efficiently a very basic low pass filter. Similar to the Gaussian kernel based interpolation this method will be denoted as S(slot width), where the slot width $W$ is again scaled by the mean sampling time.
\end{description}
Further details on each interpolation method are presented in the Supplementary Material Sec.~1. Some of these methods have been compared for classical point estimates of correlation before by \citet{rehfeld_comparison_2011}. 

\subsection{Pseudoproxies}
\label{sec:pseudoproxies}

\begin{figure}[th]

\includegraphics[width=14cm]{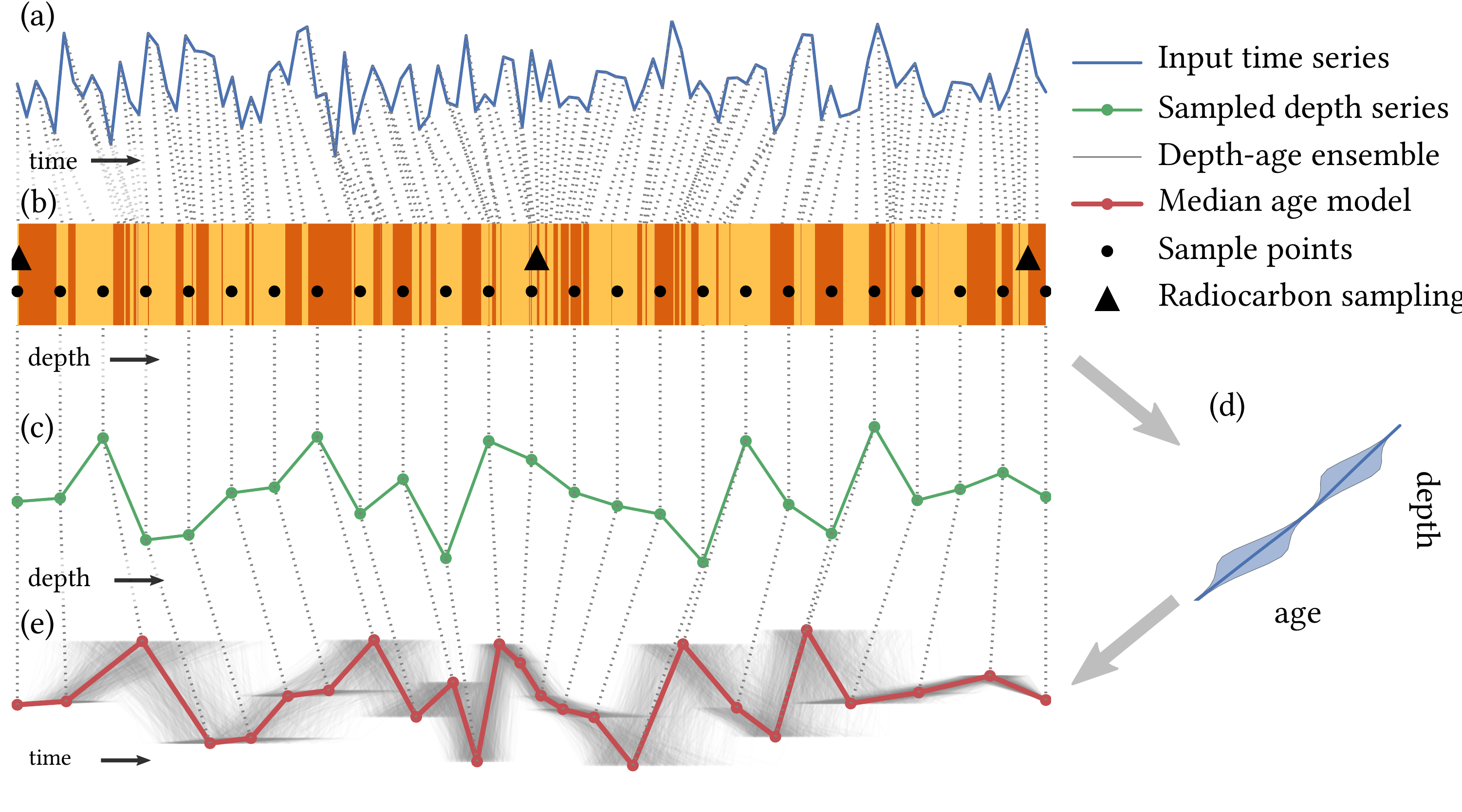}

\caption{The general procedure with which pseudoproxies are generated in this study. (a) A regularly sampled time series is generated. (b) To each value a layer of sediment is assigned, the thickness of each layer is independently drawn from a gamma distribution. (c) This sequence of layers of different thickness is sampled at regular intervals, yielding an irregularly sampled time series. (d) A small number of calendar ages are converted to radiocarbon ages and these are then used to construct a depth-age model. (e) The result is a record with a different temporal sampling due to the age model and age uncertainty.}
\label{fig:pseudoproxy}

\end{figure}

For a systematic test of how different uncertainties influence the estimation of correlation, we construct pseudoproxies resembling marine sediment records, as they incorporate many possible kinds of uncertainties present in paleoclimate proxies.

The overall procedure is sketched in Fig.~\ref{fig:pseudoproxy} and the details of the construction are presented in the Supplementary Material~Sec.~2.

The general idea is to generate a pair of coupled, regularly sampled time series for which the true correlation value is known (Fig.~\ref{fig:pseudoproxy}a). This true value is referred to as \emph{coupling strength} and denoted by $c$. Then, a sedimentation history is simulated for each proxy individually and values are assigned to layers of different width (Fig.~\ref{fig:pseudoproxy}b). Sampling at regular intervals then leads to an irregularly sampled time series of much lower resolution (Fig.~\ref{fig:pseudoproxy}c). We also simulate the effect of radiocarbon dating (Fig.~\ref{fig:pseudoproxy}d), which leads to shifted times and additional age uncertainty (Fig.~\ref{fig:pseudoproxy}e).

In our case, the pair of time series are generated using an Ornstein-Uhlenbeck process and a linearly dependent variable with a prescribed coupling strength $c$ (see Supplementary Material Sec.~S2 for details). The Ornstein-Uhlenbeck process is a continuous stochastic process which shares many characteristics of the commonly used AR(1) process, but does not rely on one discrete step-size. Therefore, it resembles the nature of proxy records more realistically, as a continuous process can be sampled at arbitrary times. The key parameter of this stochastic process is the drag parameter $\theta$ that describes the persistence structure of the time series. High values of $\theta$ correspond to predominantly random processes, while low values lead to strong persistence, common in many proxy records due to physical smoothing of the time series by diffusion, bioturbation and other mechanisms.
Accordingly, besides the coupling strength there are four more model parameters for each pair of pseudoproxy records: the length of the records, the drag parameter $\theta$ and the mean and skewness of the sedimentation rate $\mu_S$ and $\gamma_S$.

\section{Results}
\label{sec:results}

\begin{figure*}
    \begin{center}
\includegraphics[width=\textwidth]{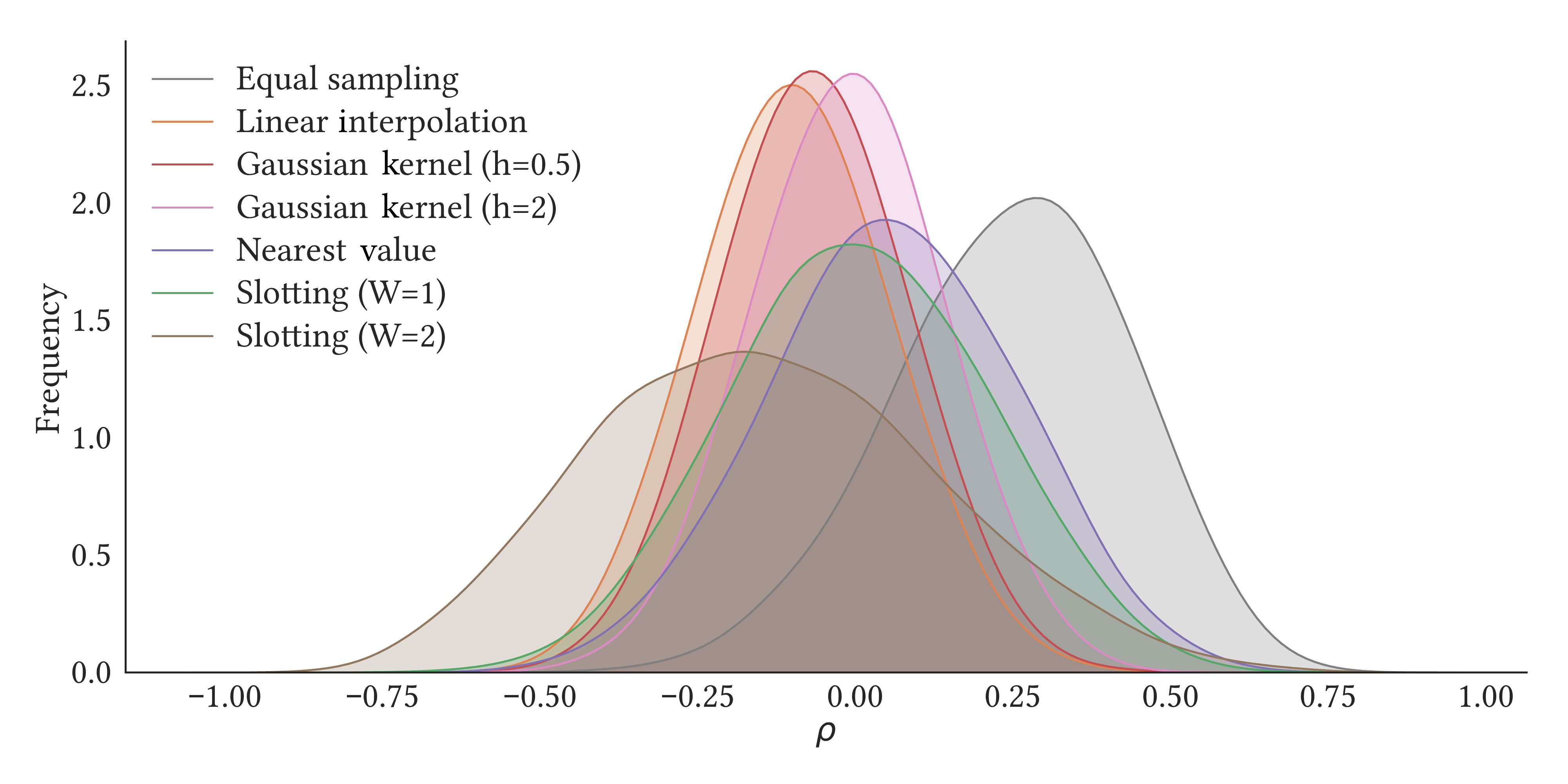}
\caption{Example for the estimation of $\rho$ using MCMC in combination with different interpolation methods. The time series in this example is a realization of the pseudoproxy described in Sec.~\ref{sec:pseudoproxies} and Supplementary Material Sec. S2 (drag parameter $\theta=0.85$). It is characterized by a true correlation value of $\rho_t = 0.3$.}
\label{fig:rhoex}
\end{center}
\end{figure*}

Before we turn to a more systematic study of correlations among pairs of pseudoproxies, a first look at an example of the estimation of $\rho_{MCMC}$ (Fig.~\ref{fig:rhoex}) is, without assuming generality, already instructive. The strength of the correlation estimated from the approximated joint distributions can be much lower than for the equally sampled case. If the values are close to zero, even different signs of the peak value $\hat{\rho}$ are possible. This demonstrates the need to make use of the full posterior distribution. Furthermore, the distribution becomes wider, and thus the estimator more uncertain, when the number of observations is reduced (as in the nearest value method or slotting with larger values of $h$).

We repeat this estimation of $\rho_{MCMC}$ for 5,000 pairs of pseudoproxies, with parameters drawn uniformly at random from the intervals shown in Tab.~\ref{tab:testvalues}. For $\theta$ and the coupling strength, we almost sample the full range of possible values, except for extreme values, which will rarely occur in nature. The values of sedimentation rate mean and skewness are inspired by the typical values of marine sediment cores that cover the late Holocene (e.g., the cores GeoB6008-1, GeoB6008-2 \citep{mcgregor_rapid_2007}, AI07-03G, AI06-04BC \citep{sicre_labrador_2014}, PL07-72GGC, PL07-73BC and CAR25-1 \citep{wurtzel_mechanisms_2013}. For each realization we test four scenarios, corresponding to different sources and degrees of uncertainties:

\begin{description}
        \item[Equal sampling.] We use the true observation times and assume that $\mathbf{Y}$ has the same sedimentation rates as $\mathbf{X}$ and thus, samples are concurrent. The main source of uncertainty is due to the finite sample size.
        \item[Unequal sampling/reference times.] We still use the true ages, but simulate separate sedimentation processes. Uncertainties thus originate from finite sample size and unequal sampling.
        \item[Age model median.] Instead of the true ages, we use the median age model $\{t^m\}$. Uncertainties include the former, while additional uncertainties come from radiocarbon calibration.
        \item[Age model ensemble.] Instead of comparing one pair of time series, we integrate over the range of realizations of the age model. This is approximated by drawing $N_{ens}$ realizations from the age model and combining the resulting posterior distributions for each ensemble member as $\rho'_{MCMC}\approx\bigcup\limits_{i=0,\dots,N_{ens}} \rho_{MCMC, i}$. This adds explicit time uncertainties. Usually, a small ensemble size of $N_{ens} \sim \mathcal{O}(10)$ is sufficient to yield a stable result.
\end{description}

Given an interpolation method $I$, a time model $T$ and a realization $i$ the resulting estimate of the correlation is denoted as $\hat{\rho}_{MCMC,i}^{I,T}$. The point-estimator for a realization $i$ is the value at which the posterior distribution of $\rho$ indicates the highest probability. The true value is considered to be $\rho = c$. The approximation methods and their corresponding parameters are summarized in Tab.~\ref{tab:interpolation} and detailed in the Supplementary Material Sec.~1.

\begin{table}
    \begin{center}
    \begin{tabular}{lll}
        parameter & abbreviation & sample interval \\
        \midrule
        Time series length & &  $[25,300]$ \\
        Coupling strength & c & $[0.1,0.9]$ \\
        Drag parameter & $\theta$ & $[0.01,0.9]$ \\
        Sedimentation rate mean & $\mu_S$ & $[0.2,0.5]$ \\
        Sedimentation rate skewness &  $\gamma_S$ & $[1,2]$
    \end{tabular}
    \caption{Intervals out of which the parameters have been drawn for the pseudoproxy experiments. Their meaning is discussed in Sec.~\ref{sec:pseudoproxies} and Supplementary Material Sec. 2. All lengths are considered to be given in a dimensionless unit. The length of a time series corresponds to the number of observations.}
    \label{tab:testvalues}
\end{center}
\end{table}

\begin{table}
    \begin{tabular}{llll}
        method & abbreviation & parameter settings \\
        \midrule
        Linear interpolation & LI & & \\
        Gaussian kernel interpolation (0.5) & G (0.5) & $h=0.5\Delta t$\\
        Gaussian kernel interpolation (2) & G (2) & $h=2\Delta t$\\
        Nearest value & NV & limit $0.5\Delta t$ \\
        Slotting (1) & S (1) & $W=\Delta t$ \\
        Slotting (2) & S (2) & $W=2\Delta t$ \\
    \end{tabular}
    \caption{Different methods to approximate the joint probability distribution in this study and the used parameters, if any. The details of these methods are discussed in Sec.~\ref{sec:irregular} and Supplementary Material Sec.~S1.}
    \label{tab:interpolation}
\end{table}

We analyze the performance of correlation estimates for different parameters of the pseudoproxy generation. To do so, we differentiate between individual series and ensemble characteristics. The former are defined on single realizations (e.g., bias), the latter on the whole ensemble of realizations (e.g., the root mean square error). 

\subsection{Individual series characteristics}
\begin{figure*}
\includegraphics[width=14cm]{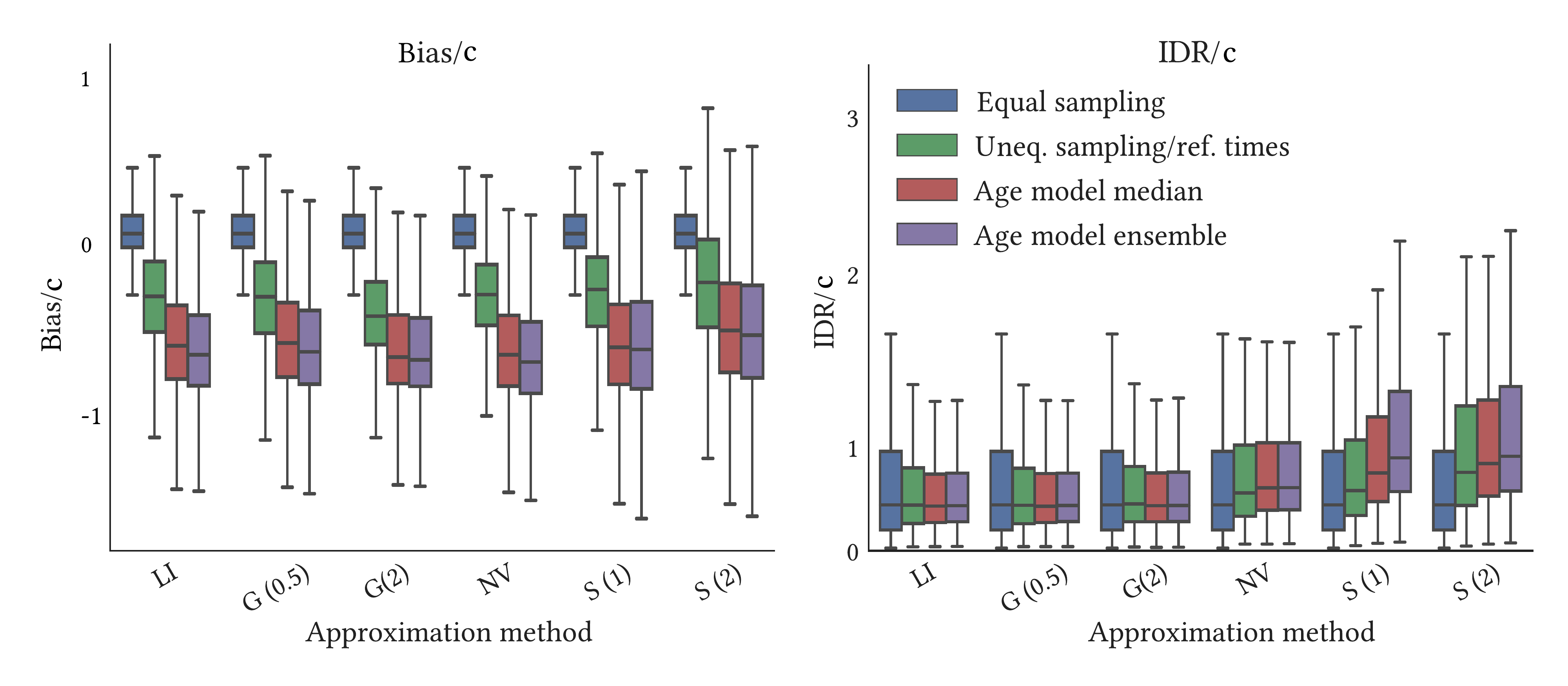}
\caption{Distribution (box plots) of scaled bias and interdecile of all realizations for different approximation methods and levels of time and sampling uncertainty. Only one value is used for each realization, the point estimate for bias and the single posterior IDR for the IDR box plot. The values are scaled to the coupling strength to make them comparable.}
\label{fig:all}
\end{figure*}

We focus on two individual series characteristics, the bias ($\text{bias}_i = \hat{\rho}_i-c$) as a measure of accuracy and the interdecile range (IDR = $(Q_{95}-Q_{5})$) as a measure of precision. The distributions of values over all realizations are shown in Fig.~\ref{fig:all}, each scaled by the true coupling strength $c$ for better comparability. In this way, a scaled bias of -1 means, that the two time series are erroneously considered independent. A similar rescaling is applied to the IDR, so that the width of the posterior distribution is related to the magnitude of the real coupling. A large IDR still yields reasonable results for large coupling strengths, but makes interpretation difficult for low coupling strengths. As the point of maximum probability is often close to the median of the posterior distribution, a scaled IDR of 0.5 and larger indicates, that a correlation of zero is likely inside the IDR.

For equally sampled time series, the median rescaled bias is very close to zero and most of the spread is due to finite sample sizes. The largest fraction of the biases is confined between -0.5 and 0.5, indicating that while finite sample sizes introduce a bias, the point-estimator still shows the sign of the correlation correctly. However, both unequal sampling and age models introduce considerable additional biases. These are mostly negative, with a tendency towards low or no correlations. Every approximation method deviates from the true variability, which can be interpreted as additional noise on the joint distribution weakening the effective coupling. The introduction of time uncertainty of the age model adds yet another bias, which is however rather small as compared to the other contributions. In general, the LI, G and NV methods show comparable results while the slotting methods exhibit larger variability but also slightly lower bias.

For the IDR we note, that the variability decreases with the introduction of additional sources of uncertainty for most approximation methods. This can be explained by the fact that interpolation methods effectively introduce new “observations“ and thus the overall set of compared values gets larger, decreasing the estimation uncertainty. Furthermore, the posterior can peak around zero if no correlation is detected, narrowing the IDR.
On the contrary, the NV and slotting methods show considerably higher IDR, which is due to the reduction of observations and, thus, higher estimation uncertainty.

To study the effects of varying parameters we focus on the realizations related to the lowest, middle and highest decile of each parameter. The corresponding results are shown in the Supplementary Material Figs.~S3 and S4. As changes in the sedimentation rate parameters $\mu_S$ and $\gamma_S$ do not result in much variation (not shown) we discuss here only the results for the effects of the time series length, coupling strength and drag parameter $\theta$.

While the median bias does not change much with increasing time series length, its variability does reduce drastically. For short time series there are more realizations (especially when using an age model) for which the scaled bias falls below $-1$, thus indicating the wrong sign of correlation. This error is markedly reduced for longer time series. A similar effect is seen for the coupling strength. While the mean scaled bias does not change much, its variability increases for weak coupling. Finally, the model parameter $\theta$ is found to have a very strong impact on the correlation estimates. For low values of $\theta$ all methods provide good estimates while for large values the introduction of unequal sampling alone leads to a scaled bias of around $-1$, so that no correlation is detected. 

To understand the latter result, we recall that the parameter $\theta$ is responsible for the persistence of the considered Ornstein-Uhlenbeck process used for generating our pseudoproxies, a low value generates strongly persistent time series and vice versa. The strength of persistence can also be estimated directly from the data, for example, via the lag-$\Delta t$ autocorrelation of $\mathbf{X}$.
We estimate the latter property with the help of the Gaussian kernel estimator \citep[denoted as gACF,][]{rehfeld_similarity_2014}. The results using the estimated persistence are very similar to those using $\theta$ alone, with almost unbiased estimates for strong persistence and very large scaled bias for weak persistence. This is not unexpected, as persistence is known to increase the values of correlation among time series in general \citep{mudelsee_climate_2010}. Also, a certain degree of persistence is necessary to be able to approximate unobserved data from close data. If there is no persistence, these close observations do not carry enough information and, hence, approximations effectively add a random signal to the time series which reduces correlation estimates.

By contrast, the IDR does not depend much on persistence. This results mainly from the fact that if no correlation is estimated, the posterior is often peaked at zero which leads to a low IDR. Besides that, we see a strong decrease in IDR for increasing time series lengths and coupling strengths. Again, the IDR is considerably larger for the nearest value and slotting methods, while linear and Gaussian kernel interpolation yield similar results. This is in line with previous discussions, e.g., by \citet{christiansen_challenges_2017}.

\subsection{Ensemble characteristics}
For most of the following results the parameter ranges are discretized to a small number of subsets and measures are then applied to these subsets. We will concentrate on two global features, the root mean square error (RMSE) and the correct sign ratio. 

\begin{figure*}
\includegraphics[width=\textwidth]{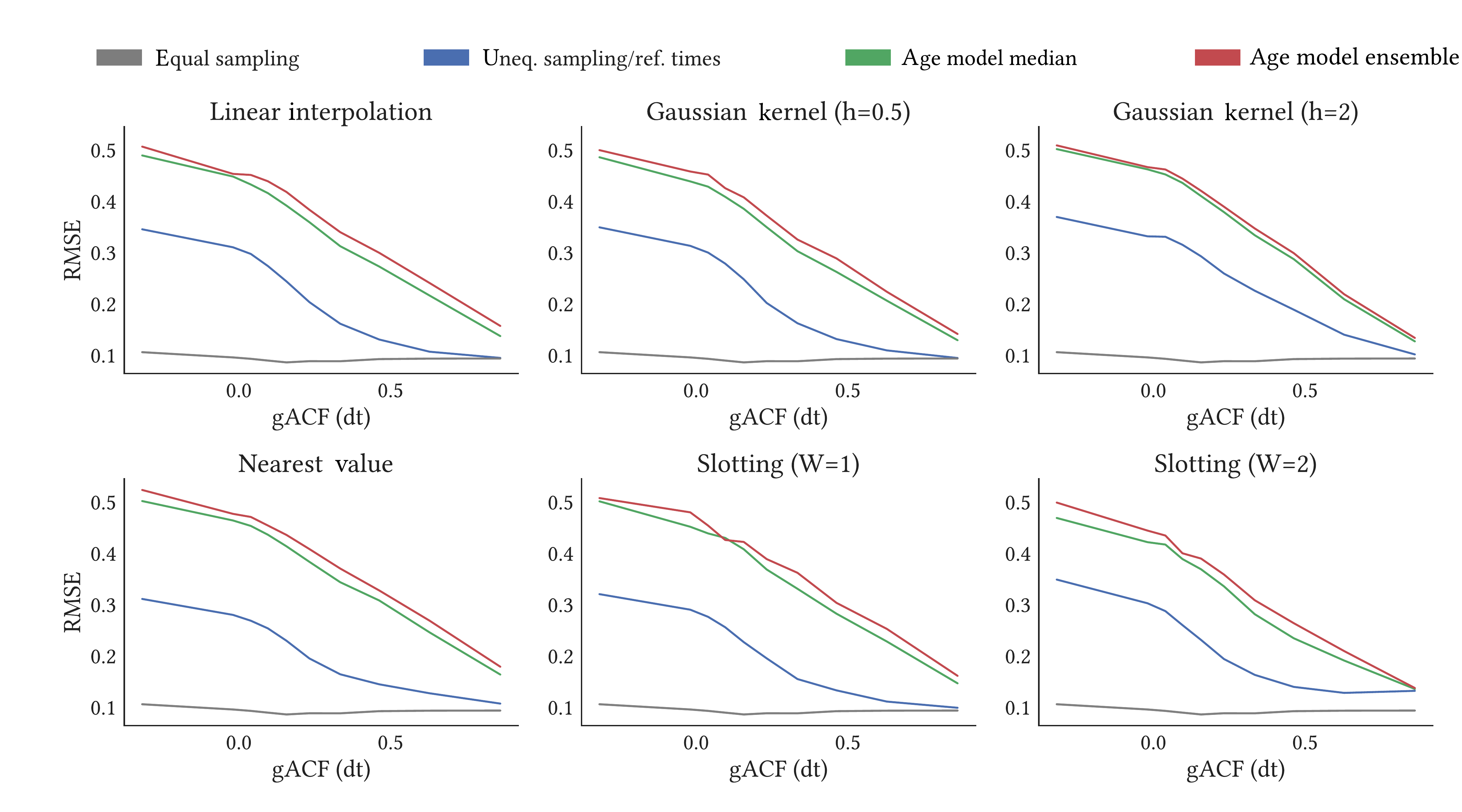}
\caption{Root mean square error (RMSE) in relation to persistence of the time series, shown for different approximation methods and different levels of time and sample uncertainty.}
\label{fig:rmse}
\end{figure*}

The RMSE is shown in Fig.~\ref{fig:rmse} in dependence on the persistence of the time series. Similar figures for the effects of time series length and coupling strength can be found in the Supplementary Material Figs.~S5 and S6.

Not surprisingly, we observe a decreasing RMSE for increasing persistence. The corresponding dependence on $\theta$ is very similar (not shown). Furthermore, there is a strong increase in RMSE with increasing coupling strength for all time models, mainly related to the increasing underestimation for large coupling strengths. Overall, the median RMSE does not decrease much at increasing sample size, even though the range of errors for single realizations that make up the RMSE is larger for short time series than for long ones, as can be seen in the Supplementary Material Fig.~S3. Again, the largest uncertainties are associated with unequal sampling and age model timescales. Additional time uncertainty does not render the observed estimates much worse. 

\begin{figure*}
    \centering
    \includegraphics[width=0.9\linewidth]{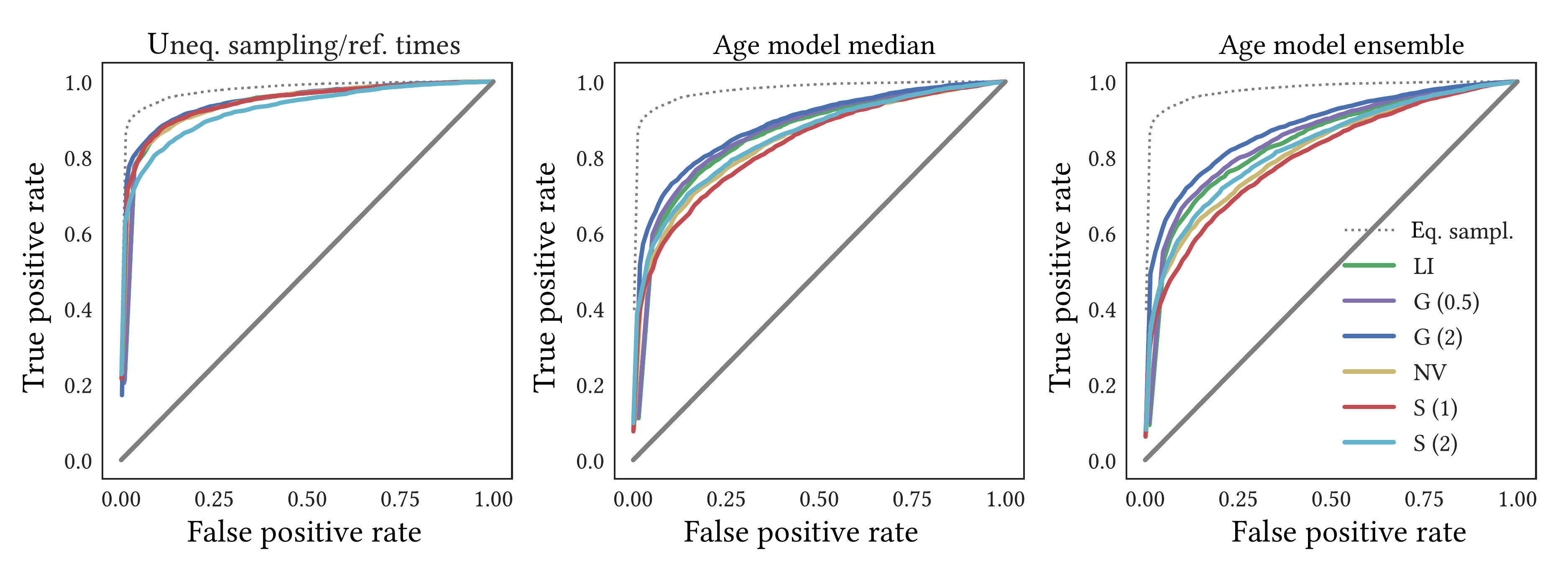}
    \caption{ROC curves for different levels of time uncertainty (from only unequal sampling on the left, to age model error in the middle and additional age model uncertainty in the right panel) and different approximation methods (shown as different colours). The dotted line is the corresponding ROC curve for equally sampled time series, where sample size is the only limitation.}
    \label{fig:roc}
\end{figure*}

In many cases, the matter of interest is not so much the exact strength of a correlation, but its presence and sign alone. From the posterior distribution $\rho_{MCMC}$ one can determine the sign by introducing a threshold level $\alpha$ and assign a positive (negative) sign  if at least a fraction $1-\alpha$ of the posterior sample values are above (below) zero. If none of the two is the case, we say that the estimator is indifferent. At first, we study how well the different methods can detect the presence or absence of correlation in general. For this, we examine the receiver operating characteristic (ROC), in which the true and false positive rates are plotted against each other for different threshold values. The resulting ROC curves are shown in Fig.~\ref{fig:roc}. While all approximation methods perform similarly well in the case of unequal sampling alone (left panel), the curves spread out when adding age model uncertainties (middle and right panel). Here, the Gaussian interpolation outperforms the other methods, in particular when using age model medians and ensembles.

\begin{figure*}
\includegraphics[width=\textwidth]{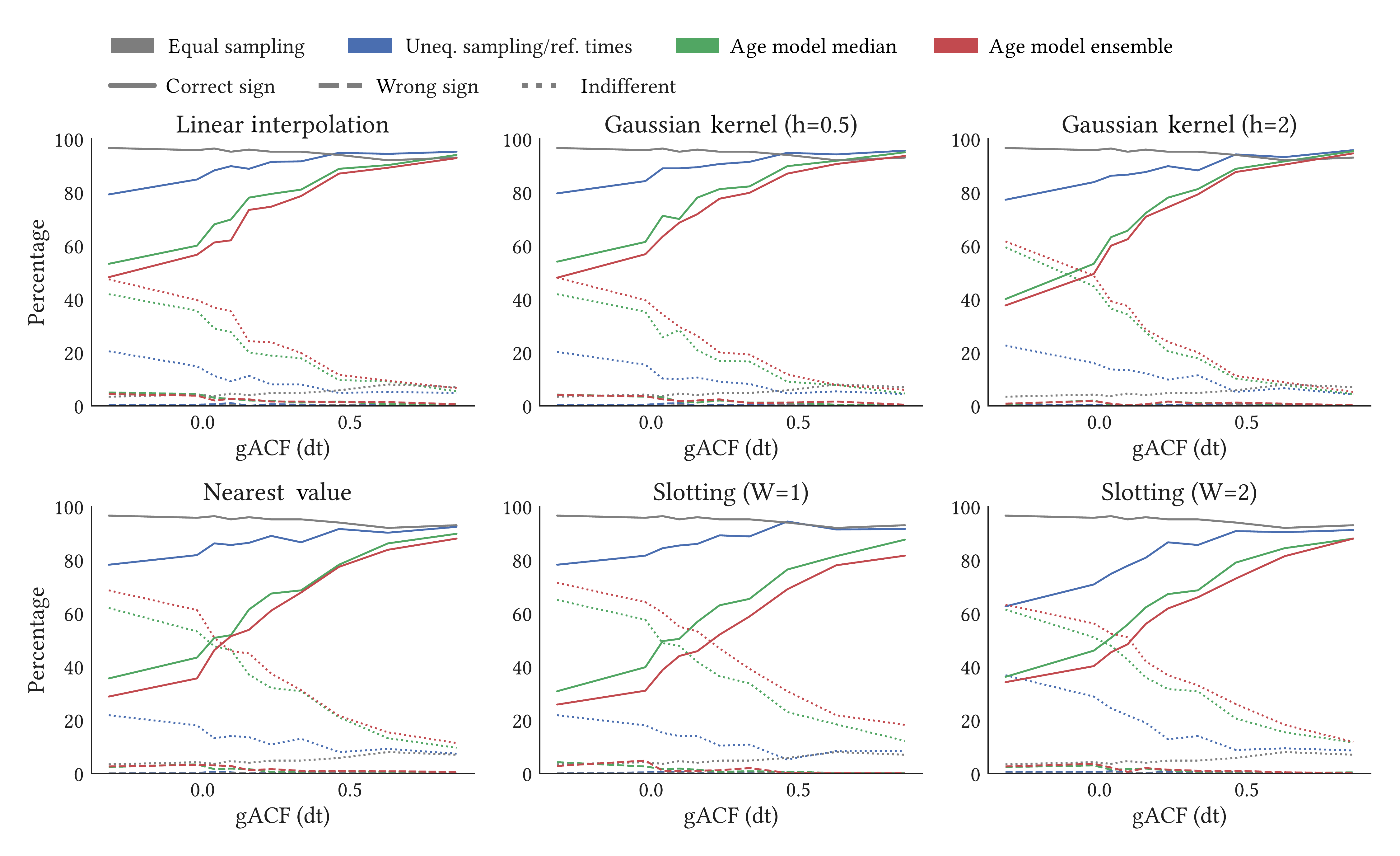}
\caption{Fractions of correctly, wrong and indifferently estimated signs of correlation in relation to persistence of the time series, shown for different approximation methods (subplots) and different levels of uncertainty (marked by colours).}
\label{fig:sign_a}
\end{figure*}

Given a subset of realizations, we can study the fraction of realizations that show a particular sign, which is shown in Fig.~\ref{fig:sign_a} and the Supplementary Material Figs.~S7 and S8. In the case of weak coupling (below 0.2) about 50\% of the estimates are indifferent, but the correlation is mostly detected successfully for stronger coupling (Supplementary Material Fig.~S7). This is even the true in $\sim 85\%$ of the cases when including age model uncertainties. 

In Fig.~\ref{fig:sign_a} and the Supplementary Material Figs.~S7 and S8 we see that the rate of correctly identified signs in the presence of age model errors and uncertainty is lower for the slotting and nearest value methods than for the interpolation methods. In the Supplementary Material Fig.~S8, a similar pattern is also seen for the dependence on the time series length (and thus effectively the sample size). Not surprisingly, the sign is determined correctly more often for large sample sizes, but there seems to be a saturation after about $200$ observations (more than present in many paleoclimate records).

Again, the persistence seems to be a good indicator of the performance of the estimator as seen in Fig.~\ref{fig:sign_a}. For persistent time series, the sign is determined correctly in most cases, independent of sampling or time uncertainties. Low or negative autocorrelation is associated with a much lower fraction of correct signs.

When it comes to the determination of the sign, most methods agree reasonably well with each other, as can be seen in the Supplementary Material Fig.~S9. Here, we show the fraction of realizations and time models for which pairs of approximation methods indicate the same sign. The closest agreement is between the LI and G (0.5) with $~91\%$. These are also the two methods which are in closest agreement for the equally sampled time series. 
In general, most of the disagreement is due to one method showing a significant correlation while another is indifferent. Disagreement on the sign of correlations is extremely rare.
The linear and Gaussian kernel interpolation seem to perform better, as the percentage of indifferent estimates is lower than for all other methods.

\section{Real world application}
\label{sec:real-world}
\begin{figure*}
\includegraphics[width=\textwidth]{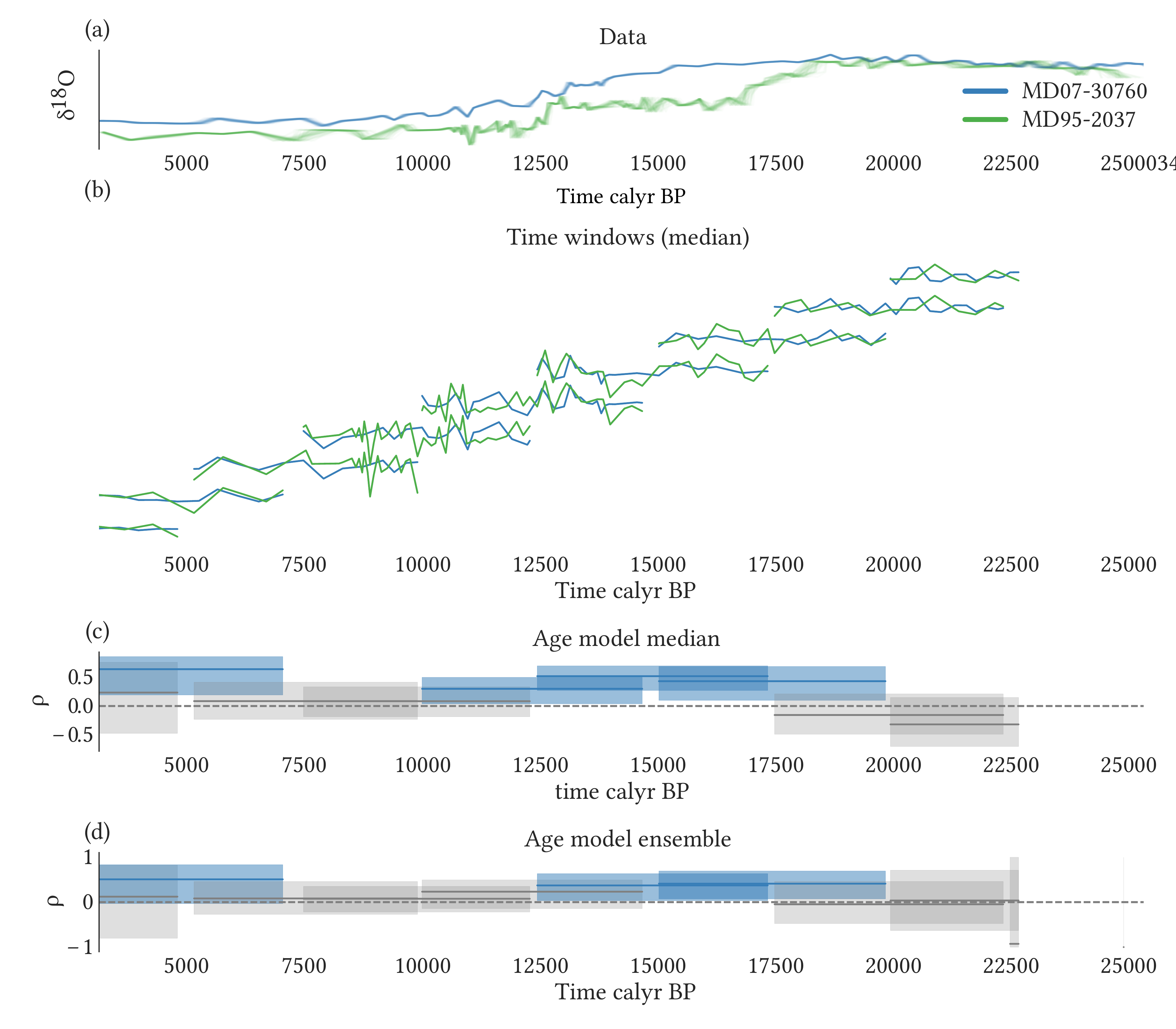}
\caption{Application of the probabilistic correlation estimation discussed in the text to a pair of marine benthic \dO records from the South (MD07-3076Q) and North Atlantic (MD95-2037). (a) Different realizations of the age model for both proxies, adjusted to a shared time axis by the time lag at which correlation is maximal. (b) Filtered values used to estimate the correlation for each window. (c) Moving window correlation when only using the median of the age model ensemble. Shown are the median value and the interdecile range. The length of each line represents the duration of the time window. Those windows for which more than 95\% of the $\rho_{MCMC}$ samples are larger or smaller than zero are shown in blue, the rest in gray. (d) Same as for (c), but for correlations integrated over a number of realizations of the age model.}
\label{fig:results_data}
\end{figure*}

To demonstrate the usefulness of a Bayesian estimation of correlation, we take a look at two marine records of benthic \dO values from the Atlantic region. The studied cores, MD07-3076Q from the South Atlantic \citep{waelbroeck_timing_2011} and MD95-2037 from the North Atlantic \citep{labeyrie_changes_2005}, both cover the last about 30 thousand years. In general, \dO is considered to be a proxy for both water temperature and salinity, but as the bottom water temperature is close to zero benthic \dO is mainly interpreted as predominantly representing salinity, which, on longer timescales, is mainly a function of the global sea ice volume \citep{bradley_paleoclimatology_2015, leng_isotopes_2006}.  

Hence, these two records are generally thought to be correlated up to a possible lag due to reservoir effects and slow mixing of the ocean water. Both are radiocarbon dated and we construct the depth-age model as used for the pseudoproxies (detailed in the Supplementary Material Sec.~1) to have comparable age uncertainties. We use our own depth-age model, because both records were published using different, and nowadays outdated, calibration curves, which show a large deviation of $\sim 400$a at the time of the deglaciation ($\sim 15$ ka BP) from the more recent calibration curve Marine13 \citep{reimer_intcal13_2013}.

MD95-2037 has a reported reservoir age effect of $\sim 400$a, while MD07-3076Q has been reported with variable reservoir effects between 1ka and 3ka. In this study, we use a relative time axis, by aligning the two records such, that the correlation is maximized, a procedure common in lead-lag estimation \citep[see e.g.][]{chang_decadal_1997, klein_remote_1999, boker_windowed_2002}. In this way, we do not include any reservoir effect into the age model. Any detected lag of the order of 1 to 3 ka could thus result from such reservoir effects.

For simplicity, we use linear interpolation in this application. We find a lag of 2100 years to yield a maximum correlation value. Nevertheless, the correlation posteriors are very similar for all lag values between 1.5ka and 3.1ka and thus, the lag relationship shows a considerable degree of uncertainty. 

In Fig.~\ref{fig:results_data}a one can see different realizations of the age models of the two records, adjusted for the mutual lag showing the maximum correlation. Even though both records exhibit a marked drop in \dO during the deglaciation, the shape of the transition is considerably different. Also, it is difficult to assess the similarity at shorter timescales. \citet{waelbroeck_timing_2011} discussed, that benthic \dO records might not be correlated at sub-millennial timescales. 

To concentrate on different time-scales of interest we apply Complete Ensemble Empirical Mode Decomposition with Adaptive Noise (CEEMDAN, \citet{torres_complete_2011}) to each time series and discard those intrinsic mode functions (IMFs) for which more than 10\% of the instantaneous wavelengths are larger than 5ka. This ensures, that no overlying trends and low-frequency variability alter the results obtained in the following.

For the time series reconstructed by superposition of the higher-frequency IMFs, the maximum correlation for the age model median and the age model ensemble are reached at lags of 2.1ka and 2.05ka, respectively, comparable with the results for the original unfiltered records.

Having adjusted the two decomposed time series on a shared timescale, we now use sliding windows of 5ka length and mutual overlap of 2.5ka to study changes in covariability over time. The results are shown in Figs.~\ref{fig:results_data}c, d. 

For the age model median, we see significant correlation over most of the last 22 millennia. Only at the beginning of the Holocene and towards the end of the records (probably related to a lower number of samples) there are no significant correlations. If we take age uncertainty into account we generally find a similar pattern, but the estimator is now indifferent for one more  time window. 

\section{Discussion and conclusions}

The aim of this study was to provide some guidelines for the estimation of correlations between (marine) paleoclimate records, in particular when dealing with unequal sampling and age model uncertainties. The main tool has been the probabilistic (Bayesian) estimation based on a bivariate normal model. While most results might not be surprising, they have rarely been addressed explicitly in previous works. 

In cases of weak coupling, different approximation methods can show even different signs of the estimated correlation, but these are rarely significant. Thus, when different approximation methods with different parameters yield different signs for the point-estimator it is crucial to take estimation uncertainty into account. This will probably lead to a non-significant correlation.
Still, in most cases, the correct sign of a correlation can be detected even in the presence of a large bias of its absolute value.

Those methods which reduce the number of observations like slotting tend to highlight low-frequency trends. In comparison, interpolation seems to be a better strategy to approximate the joint probability distribution. The best results were obtained using linear interpolation or a Gaussian kernel with small bandwidth. For small bandwidths, the directly neighboring observations dominate the approximation of missing values and thus these two methods are expected to be very similar. 

Our findings agree with other studies, that have reported the Gaussian kernel estimator to be superior \citep{rehfeld_comparison_2011, rehfeld_similarity_2014}, even though the difference to linear interpolation is small. Hence, it seems reasonable to prefer the Gaussian kernel estimator with small bandwidth over other methods. These previous studies have used classical point estimates, e.g., Pearson's linear correlation coefficient. In this study, we have thus shown that the Bayesian estimation leads to comparable results, while offering additional advantages as a straightforward treatment of uncertainties and well-defined correlation matrices.

The characteristics of the time series under study are in many cases more important than the choice of the particular interpolation method. The most important features are the coupling strength, number of observations, and persistence of the signal. 

The first factor is generally not accessible for real-world data, but the other two can be estimated. As mentioned in Sec.~\ref{sec:results} persistence has been seen as a problem to correlation estimation due to its tendency to induce spuriously high correlations. Still, it is of advantage in the case of irregularly sampled time series, as it allows inference from neighbouring points on unobserved data. As persistence is necessary to estimate correlations for unevenly sampled time series, it would be beneficial to also include it in the statistical model in the first place. This could be done by moving away from the bivariate normal random variables model towards more general bivariate stochastic processes which additionally include autocorrelations. However, strongly irregular sampling would make it difficult to use common models of persistence like autoregressive processes, as they can only be approximated  (similar to the joint distributions) and restrict the persistence to one time scale (as given by the fixed sampling interval of such processes). This restriction does not seem to be appropriate for many paleoclimate records. While a detailed study of appropriate persistence models should be performed in future work, the results presented here stress the need for such models.

Thus, in real-world applications, one should check, if a time series shows enough persistence to perform meaningful interpolation. While the sample length does not affect the error much, it increases the estimation uncertainty. Most time uncertainties mainly lead to indifferent estimates, but rarely produce false signs of correlation. If one finds a significant correlation between unequally and age-uncertain time series one can hence accept them to reflect features of the time series. Large persistence nevertheless yields the same problems as in the case of regularly and evenly sampled time series. Thus, if a significant correlation is found, it should be investigated if it might be due to low-frequency variability (if the latter is not the time-scale of interest).

In most of the pseudoproxy experiments in this study, the additional error from time uncertainty has been much smaller than those originating from unequal sampling and age model calibration. From our real-world example we can see that many features are similar in both cases, even though sometimes significance is lost. In the past, only few authors have published explicit age uncertainties for each observation. In these situations, it might still be preferable to use the published age model, as a bad age model with quantified time uncertainties might introduce a larger error than discarding time uncertainties. 

In this study, we have limited ourselves to marine sediment records, as we consider them to be more extremely affected by these uncertainties than many other archives of Quaternary climate variability like tree rings, corals, lake sediments, ice cores and speleothems. Some of the aforementioned types of archives commonly have relatively reliable age measurements (e.g., based on layer counts in ice cores, varved lake sediments, or corals) so that we can expect the effects of age error to be considerably smaller. In turn, non-varved lake sediments or speleothem records may suffer from similar effects like marine sediments, and are therefore not systematically considered in the following. A more detailed and systematic discussion of differences in age uncertainties in different archives, possibly using multiple forward proxy models, has been beyond the scope of the present paper.

The uncertainty in observation times is rather small in our age model as compared to those produced by other age model algorithms like Bacon \citep{blaauw_flexible_2011}. Hence, when using such models it is possible that fewer time intervals will show significant correlations. We thus decided against using these more elaborate age model programs, as this would make comparison with the previous results difficult. A wider posterior distribution in the sampling intervals broadens the posterior distributions of the correlation strength and hence lead to more false negatives in the case of more elaborate age models and more false positives for simple models, e.g., those based on linear interpolation between age control points. In terms of complexity, we consider our age model to sit in between rather simple (e.g. clam) and complex age models (like Bacon or Bchron). Therefore, we believe that our results can be related to applications using other age model algorithms as well.

We have finally discussed an example of two benthic \dO records, which are generally thought to be correlated on at least millennial timescales. Using the Bayesian estimation of correlation together with Gaussian kernel based interpolation we have seen that when adjusting to one time-scale we find similarities even at the sub-millennial scales. These correlations are not stationary and change over time, being significant only during some periods. Taking uncertainties into account can be helpful in assessing if two time series are similar to each other. In summary, we therefore conclude that it would be useful if data publishers would also report the age uncertainty given by their age model to each observation to make more accurate estimations possible.

\section*{Acknowledgements}
This work has been financially supported by the German Federal Ministry for Education and Research (BMBF) via the BMBF Young Investigators Group "CoSy-CC$^2$ - Complex Systems Approaches to Understanding Causes and Consequences of Past, Present and Future Climate Change" (grant no. 01LN1306A), and by the joint German-Norwegian project "Nonlinear variability and regime shifts in Late Holocene climate: regional patterns and inter-regional linkages in multi-proxy networks and climate simulations" jointly funded by the German Academic Exchange Service (DAAD project no. 57245873) and the Research Council of Norway. The authors thank Johannes Werner and Dimitry Divine for fruitful discussions.

\section*{Code availability}
A small python script to sample the posterior distribution using the pyMC3 package can be found at
\url{https://github.com/jzzbr/bayesian_correlation}.


  \bibliographystyle{elsarticle-harv} 
  \bibliography{bibliography_franke_corr.bib}

\begin{thebibliography}{}

\bibitem[\protect\astroncite{Babu and Stoica}{2010}]{babu_spectral_2010}
Babu, P. and P.~Stoica\leavevmode\nopagebreak\newline 2010.
\newblock Spectral analysis of nonuniformly sampled data \textendash{} a
  review.
\newblock {\em Digital Signal Processing}, 20(2):359--378.

\bibitem[\protect\astroncite{Behseta et~al.}{2009}]{behseta_bayesian_2009}
Behseta, S., T.~Berdyyeva, C.~R. Olson, and R.~E.
  Kass\leavevmode\nopagebreak\newline 2009.
\newblock Bayesian {{Correction}} for {{Attenuation}} of {{Correlation}} in
  {{Multi}}-{{Trial Spike Count Data}}.
\newblock {\em Journal of Neurophysiology}, 101(4):2186--2193.

\bibitem[\protect\astroncite{Blaauw}{2010}]{blaauw_methods_2010}
Blaauw, M.\leavevmode\nopagebreak\newline 2010.
\newblock Methods and code for `classical' age-modelling of radiocarbon
  sequences.
\newblock {\em Quaternary Geochronology}, 5(5):512--518.

\bibitem[\protect\astroncite{Blaauw and Christen}{2011}]{blaauw_flexible_2011}
Blaauw, M. and J.~A. Christen\leavevmode\nopagebreak\newline 2011.
\newblock Flexible {{Paleoclimate Age}}-{{Depth Models Using}} an
  {{Autoregressive Gamma Process}}.
\newblock {\em Bayesian Analysis}, 6(3):457--474.

\bibitem[\protect\astroncite{Boker et~al.}{2002}]{boker_windowed_2002}
Boker, S.~M., J.~L. Rotondo, M.~Xu, and K.~King\leavevmode\nopagebreak\newline
  2002.
\newblock Windowed cross-correlation and peak picking for the analysis of
  variability in the association between behavioral time series.
\newblock {\em Psychological Methods}, 7(3):338--355.

\bibitem[\protect\astroncite{Bradley}{2015}]{bradley_paleoclimatology_2015}
Bradley, R.~S.\leavevmode\nopagebreak\newline 2015.
\newblock {\em Paleoclimatology: Reconstructing Climates of the Quaternary}, 3.
  ed edition.
\newblock Amsterdam: {Academic Press/Elsevier}.

\bibitem[\protect\astroncite{Chang et~al.}{1997}]{chang_decadal_1997}
Chang, P., L.~Ji, and H.~Li\leavevmode\nopagebreak\newline 1997.
\newblock A decadal climate variation in the tropical {{Atlantic Ocean}} from
  thermodynamic air-sea interactions.
\newblock {\em Nature}, 385(6616):516--518.

\bibitem[\protect\astroncite{Cheng et~al.}{2012}]{cheng_climatic_2012}
Cheng, H., P.~Z. Zhang, C.~Sp\"otl, R.~L. Edwards, Y.~J. Cai, D.~Z. Zhang,
  W.~C. Sang, M.~Tan, and Z.~S. An\leavevmode\nopagebreak\newline 2012.
\newblock The climatic cyclicity in semiarid-arid central {{Asia}} over the
  past 500,000 years.
\newblock {\em Geophysical Research Letters}, 39(1):L01705.

\bibitem[\protect\astroncite{Christiansen and
  Ljungqvist}{2017}]{christiansen_challenges_2017}
Christiansen, B. and F.~C. Ljungqvist\leavevmode\nopagebreak\newline 2017.
\newblock Challenges and {{Perspectives}} for {{Large}}-{{Scale Temperature
  Reconstructions}} of the {{Past Two Millennia}}.
\newblock {\em Reviews of Geophysics}, 55(1):40--96.

\bibitem[\protect\astroncite{Fisher}{1915}]{fisher_frequency_1915}
Fisher, R.~A.\leavevmode\nopagebreak\newline 1915.
\newblock Frequency {{Distribution}} of the {{Values}} of the {{Correlation
  Coefficient}} in {{Samples}} from an {{Indefinitely Large Population}}.
\newblock {\em Biometrika}, 10(4):507--521.

\bibitem[\protect\astroncite{Franke et~al.}{2017}]{nao_paper}
Franke, J.~G., J.~P. Werner, and R.~V. Donner\leavevmode\nopagebreak\newline
  2017.
\newblock Reconstructing {{Late Holocene North Atlantic}} atmospheric
  circulation changes using functional paleoclimate networks.
\newblock {\em Climate of the Past}, 13(11):1593--1608.

\bibitem[\protect\astroncite{Gelman}{2014}]{gelman_bayesian_2014}
Gelman, A.\leavevmode\nopagebreak\newline 2014.
\newblock {\em Bayesian Data Analysis}, 3rd edition.
\newblock Boca Raton: {CRC Press}.

\bibitem[\protect\astroncite{Klein et~al.}{1999}]{klein_remote_1999}
Klein, S.~A., B.~J. Soden, and N.-C. Lau\leavevmode\nopagebreak\newline 1999.
\newblock Remote {{Sea Surface Temperature Variations}} during {{ENSO}}:
  {{Evidence}} for a {{Tropical Atmospheric Bridge}}.
\newblock {\em Journal of Climate}, 12(4):917--932.

\bibitem[\protect\astroncite{Labeyrie et~al.}{2005}]{labeyrie_changes_2005}
Labeyrie, L., C.~Waelbroeck, E.~Cortijo, E.~Michel, and J.-C.
  Duplessy\leavevmode\nopagebreak\newline 2005.
\newblock Changes in deep water hydrology during the {{Last Deglaciation}}.
\newblock {\em Comptes Rendus Geoscience}, 337(10-11):919--927.

\bibitem[\protect\astroncite{Lehmann and Casella}{1998}]{lehmann_theory_1998}
Lehmann, E.~L. and G.~Casella\leavevmode\nopagebreak\newline 1998.
\newblock {\em Theory of Point Estimation}, 2nd edition.
\newblock New York: {Springer}.

\bibitem[\protect\astroncite{Lewandowski
  et~al.}{2009}]{lewandowski_generating_2009}
Lewandowski, D., D.~Kurowicka, and H.~Joe\leavevmode\nopagebreak\newline 2009.
\newblock Generating random correlation matrices based on vines and extended
  onion method.
\newblock {\em Journal of Multivariate Analysis}, 100(9):1989--2001.

\bibitem[\protect\astroncite{Mann et~al.}{1998}]{mann_global-scale_1998}
Mann, M.~E., R.~S. Bradley, and M.~K. Hughes\leavevmode\nopagebreak\newline
  1998.
\newblock Global-scale temperature patterns and climate forcing over the past
  six centuries.
\newblock {\em Nature}, 392:779--787.

\bibitem[\protect\astroncite{Marcott
  et~al.}{2013}]{marcott_reconstruction_2013}
Marcott, S.~A., J.~D. Shakun, P.~U. Clark, and A.~C.
  Mix\leavevmode\nopagebreak\newline 2013.
\newblock A {{Reconstruction}} of {{Regional}} and {{Global Temperature}} for
  the {{Past}} 11,300 {{Years}}.
\newblock {\em Science}, 339(6124):1198--1201.

\bibitem[\protect\astroncite{Maslin and Swann}{2006}]{leng_isotopes_2006}
Maslin, M.~A. and G.~E. Swann\leavevmode\nopagebreak\newline 2006.
\newblock Isotopes in marine sediments.
\newblock In {\em Isotopes in {{Palaeoenvironmental Research}}}, M.~J. Leng,
  ed., volume~10, Pp.~ 227--290.
\newblock Dordrecht: {Kluwer Academic Publishers}.

\bibitem[\protect\astroncite{Matzke et~al.}{2017}]{matzke_bayesian_2017}
Matzke, D., A.~Ly, R.~Selker, W.~D. Weeda, B.~Scheibehenne, M.~D. Lee, and
  E.-J. Wagenmakers\leavevmode\nopagebreak\newline 2017.
\newblock Bayesian {{Inference}} for {{Correlations}} in the {{Presence}} of
  {{Measurement Error}} and {{Estimation Uncertainty}}.
\newblock {\em Collabra: Psychology}, 3(1):25.

\bibitem[\protect\astroncite{McGregor et~al.}{2007}]{mcgregor_rapid_2007}
McGregor, H.~V., M.~Dima, H.~W. Fischer, and
  S.~Mulitza\leavevmode\nopagebreak\newline 2007.
\newblock Rapid 20th-{{Century Increase}} in {{Coastal Upwelling}} off
  {{Northwest Africa}}.
\newblock {\em Science}, 315(5812):637--639.

\bibitem[\protect\astroncite{Mudelsee}{2010}]{mudelsee_climate_2010}
Mudelsee, M.\leavevmode\nopagebreak\newline 2010.
\newblock {\em Climate {{Time Series Analysis}}}, volume~42 of {\em Atmospheric
  and {{Oceanographic Sciences Library}}}.
\newblock Dordrecht: {Springer Netherlands}.

\bibitem[\protect\astroncite{{PAGES 2k
  Consortium}}{2013}]{pages_2k_consortium_continental-scale_2013}
{PAGES 2k Consortium}\leavevmode\nopagebreak\newline 2013.
\newblock Continental-scale temperature variability during the past two
  millennia.
\newblock {\em Nature Geoscience}, 6(5):339--346.

\bibitem[\protect\astroncite{Paninski}{2003}]{paninski_estimation_2003}
Paninski, L.\leavevmode\nopagebreak\newline 2003.
\newblock Estimation of {{Entropy}} and {{Mutual Information}}.
\newblock {\em Neural Computation}, 15(6):1191--1253.

\bibitem[\protect\astroncite{Papana and
  Kugiumtzis}{2009}]{papana_evaluation_2009}
Papana, A. and D.~Kugiumtzis\leavevmode\nopagebreak\newline 2009.
\newblock Evaluation of mutual information estimators for time series.
\newblock {\em International Journal of Bifurcation and Chaos},
  19(12):4197--4215.

\bibitem[\protect\astroncite{Polson and Scott}{2012}]{polson_half-cauchy_2012}
Polson, N.~G. and J.~G. Scott\leavevmode\nopagebreak\newline 2012.
\newblock On the {{Half}}-{{Cauchy Prior}} for a {{Global Scale Parameter}}.
\newblock {\em Bayesian Analysis}, 7(4):887--902.

\bibitem[\protect\astroncite{Rehfeld and
  Kurths}{2014}]{rehfeld_similarity_2014}
Rehfeld, K. and J.~Kurths\leavevmode\nopagebreak\newline 2014.
\newblock Similarity {{Estimators}} for {{Irregular}} and {{Age}}-{{Uncertain
  Time Series}}.
\newblock {\em Climate of the Past}, 10(1):107--122.

\bibitem[\protect\astroncite{Rehfeld et~al.}{2013}]{rehfeld_late_2013}
Rehfeld, K., N.~Marwan, S.~F.~M. Breitenbach, and
  J.~Kurths\leavevmode\nopagebreak\newline 2013.
\newblock Late {{Holocene Asian}} summer monsoon dynamics from small but
  complex networks of paleoclimate data.
\newblock {\em Climate Dynamics}, 41(1):3--19.

\bibitem[\protect\astroncite{Rehfeld et~al.}{2011}]{rehfeld_comparison_2011}
Rehfeld, K., N.~Marwan, J.~Heitzig, and
  J.~Kurths\leavevmode\nopagebreak\newline 2011.
\newblock Comparison of {{Correlation Analysis Techniques}} for {{Irregularly
  Sampled Time Series}}.
\newblock {\em Nonlinear Processes in Geophysics}, 18(3):389--404.

\bibitem[\protect\astroncite{Reimer et~al.}{2013}]{reimer_intcal13_2013}
Reimer, P.~J., E.~Bard, A.~Bayliss, J.~W. Beck, P.~G. Blackwell, C.~B. Ramsey,
  C.~E. Buck, H.~Cheng, R.~L. Edwards, M.~Friedrich, P.~M. Grootes, T.~P.
  Guilderson, H.~Haflidason, I.~Hajdas, C.~Hatt\'e, T.~J. Heaton, D.~L.
  Hoffmann, A.~G. Hogg, K.~A. Hughen, K.~F. Kaiser, B.~Kromer, S.~W. Manning,
  M.~Niu, R.~W. Reimer, D.~A. Richards, E.~M. Scott, J.~R. Southon, R.~A.
  Staff, C.~S.~M. Turney, and J.~{van der
  Plicht}\leavevmode\nopagebreak\newline 2013.
\newblock {{IntCal13}} and {{Marine13 Radiocarbon Age Calibration Curves}}
  0\textendash{}50,000 {{Years}} cal {{BP}}.
\newblock {\em Radiocarbon}, 55(04):1869--1887.

\bibitem[\protect\astroncite{Sicre et~al.}{2014}]{sicre_labrador_2014}
Sicre, M.-A., K.~Weckstr\"om, M.-S. Seidenkrantz, A.~Kuijpers, M.~Benetti,
  G.~Masse, U.~Ezat, S.~Schmidt, I.~Bouloubassi, J.~Olsen, M.~Khodri, and
  J.~Mignot\leavevmode\nopagebreak\newline 2014.
\newblock Labrador current variability over the last 2000 years.
\newblock {\em Earth and Planetary Science Letters}, 400:26--32.

\bibitem[\protect\astroncite{Smerdon and
  Pollack}{2016}]{smerdon_reconstructing_2016}
Smerdon, J.~E. and H.~N. Pollack\leavevmode\nopagebreak\newline 2016.
\newblock Reconstructing {{Earth}}'s surface temperature over the past 2000
  years: The science behind the headlines: {{Reconstructing Earth}}'s surface
  temperature over the past 2000 years.
\newblock {\em Wiley Interdisciplinary Reviews: Climate Change}, 7(5):746--771.

\bibitem[\protect\astroncite{Spearman}{1904}]{spearman_proof_1904}
Spearman, C.\leavevmode\nopagebreak\newline 1904.
\newblock The {{Proof}} and {{Measurement}} of {{Association}} between {{Two
  Things}}.
\newblock {\em The American Journal of Psychology}, 15(1):72--101.

\bibitem[\protect\astroncite{Tingley et~al.}{2012}]{tingley_piecing_2012}
Tingley, M.~P., P.~F. Craigmile, M.~Haran, B.~Li, E.~Mannshardt, and
  B.~Rajaratnam\leavevmode\nopagebreak\newline 2012.
\newblock Piecing together the past: Statistical insights into paleoclimatic
  reconstructions.
\newblock {\em Quaternary Science Reviews}, 35:1--22.

\bibitem[\protect\astroncite{Torres et~al.}{2011}]{torres_complete_2011}
Torres, M.~E., M.~A. Colominas, G.~Schlotthauer, and
  P.~Flandrin\leavevmode\nopagebreak\newline 2011.
\newblock A complete ensemble empirical mode decomposition with adaptive noise.
\newblock Pp.~ 4144--4147. {IEEE}.

\bibitem[\protect\astroncite{{von
  Toussaint}}{2011}]{von_toussaint_bayesian_2011}
{von Toussaint}, U.\leavevmode\nopagebreak\newline 2011.
\newblock Bayesian inference in physics.
\newblock {\em Reviews of Modern Physics}, 83(3):943--999.

\bibitem[\protect\astroncite{Waelbroeck et~al.}{2011}]{waelbroeck_timing_2011}
Waelbroeck, C., L.~C. Skinner, L.~Labeyrie, J.-C. Duplessy, E.~Michel,
  N.~Vazquez~Riveiros, J.-M. Gherardi, and
  F.~Dewilde\leavevmode\nopagebreak\newline 2011.
\newblock The timing of deglacial circulation changes in the {{Atlantic}}.
\newblock {\em Paleoceanography}, 26(3):PA3213.

\bibitem[\protect\astroncite{Werner et~al.}{2018}]{werner_spatio-temporal_2018}
Werner, J.~P., D.~V. Divine, F.~Charpentier~Ljungqvist, T.~Nilsen, and
  P.~Francus\leavevmode\nopagebreak\newline 2018.
\newblock Spatio-temporal variability of {{Arctic}} summer temperatures over
  the past 2 millennia.
\newblock {\em Climate of the Past}, 14(4):527--557.

\bibitem[\protect\astroncite{Wurtzel et~al.}{2013}]{wurtzel_mechanisms_2013}
Wurtzel, J.~B., D.~E. Black, R.~C. Thunell, L.~C. Peterson, E.~J. Tappa, and
  S.~Rahman\leavevmode\nopagebreak\newline 2013.
\newblock Mechanisms of southern {{Caribbean SST}} variability over the last
  two millennia: {{ATLANTIC SSTs OVER THE LAST}} 2000 {{YEARS}}.
\newblock {\em Geophysical Research Letters}, 40(22):5954--5958.

\bibitem[\protect\astroncite{Zhang et~al.}{2008}]{zhang_test_2008}
Zhang, P., H.~Cheng, R.~L. Edwards, F.~Chen, Y.~Wang, X.~Yang, J.~Liu, M.~Tan,
  X.~Wang, J.~Liu, C.~An, Z.~Dai, J.~Zhou, D.~Zhang, J.~Jia, L.~Jin, and K.~R.
  Johnson\leavevmode\nopagebreak\newline 2008.
\newblock A {{Test}} of {{Climate}}, {{Sun}}, and {{Culture Relationships}}
  from an 1810-{{Year Chinese Cave Record}}.
\newblock {\em Science}, 322(5903):940--942.

\end{thebibliography}


\end{document}